\documentclass[preprint2,times]{aastex62}
\usepackage{apjfonts}
\usepackage{enumitem}
\usepackage{color}

\newcommand{\eqref}[1]{Equation (\ref{#1})}

\shorttitle{Normal Double Detonation SNe~Ia}

\begin{document}

\title{Double Detonations with Thin, Modestly Enriched Helium Layers Can Make Normal Type Ia Supernovae}

\author[0000-0002-9538-5948]{Dean M. Townsley}
\affiliation{Department of Physics \& Astronomy,
	University of Alabama, Tuscaloosa, AL, USA, Email:Dean.M.Townsley@ua.edu}
\author{Broxton J. Miles}
\affiliation{Department of Physics, North Carolina State University, Raleigh, NC, USA}
\author[0000-0002-9632-6106]{Ken J. Shen}
\affiliation{Department of Astronomy and Theoretical Astrophysics Center,
	University of California, Berkeley, CA 94720, USA}
\author{Daniel Kasen}
\affiliation{Department of Astronomy and Theoretical Astrophysics Center,
	University of California, Berkeley, CA 94720, USA}

\begin{abstract}

It has been proposed that Type Ia supernovae (SNe Ia) that are normal in their spectra and brightness can be explained by a double detonation that ignites first in a helium shell on the surface of the white dwarf (WD).
This proposition is supported by the satisfactory match between simulated explosions of sub-Chandrasekhar-mass WDs with no surface He layer and observations of normal SNe Ia.
However, previous calculations of He-ignited double detonations have required either artificial removal of the He shell ashes or extreme enrichment of the surface He layer in order to obtain normal SNe Ia.
Here we demonstrate, for the first time in multi-dimensional full-star simulations, that a thin, modestly enriched He layer will lead to a SN Ia that is normal in its brightness and spectra.
This strengthens the case for double detonations as a major contributing channel to the population of normal SNe Ia.

\end{abstract}


\section{Context and Choice of Progenitor}
\label{sec:intro}

There is broad consensus that Type Ia supernovae (SNe~Ia), whose spectra lack H and He features but contain characteristic Si features, are produced by the incineration of a carbon-oxygen-rich white dwarf (WD) star
(see \citealt{SeitenzahlTownsley2017} and \citealt{RopkeSim2018} for recent reviews),
with radioactive decay of nickel providing the power for their optically bright phase.
However, it remains challenging to construct a scenario for spectroscopically normal \citep{Branchetal2006,Parrentetal2014} SNe~Ia that satisfies constraints posed by both the large body of observations available and the extensive study of the physical processes involved.

While buildup of a WD toward the Chandrasekhar limiting mass provides a fairly straightforward mechanism for the ignition of fusion, Chandrasekhar-mass white dwarfs require some form of expansion before their full incineration in order to reproduce normal SNe~Ia
\citep{Nomotoetal1984,Khokhlov1991}.
A sub-Chandrasekhar mass white dwarf incinerated by a detonation will make a balance of elements similar to that observed in SNe~Ia \citep{Simetal2010,Shen_18}.
One possible ignition scenario for a sub-Chandrasekhar mass WD begins in a helium shell which then triggers the detonation of the C-O core: a double detonation.
As initially proposed \citep{Taam1980,Nomoto1980,nomo82b,wtw86}, the detonation proceeded directly from the He shell into the core.
Such a direct transition of the detonation from He to C-O was found to be more difficult than initially thought \citep{LivneGlasner1990},
but it was also found that the He detonation should make an inward shock wave that will focus and compress material in the core, leading to a successive C-O ignition and explosion \citep{Livne1990,LivneGlasner1991,LivneArnett1995}.

While promising, further calculations bore out concerns that this scenario, with He shells around 0.1~M$_\odot$, did not reproduce normal SNe~Ia in decline rate \citep{WoosleyWeaver1994,HoeflichKhokhlov1996} or spectroscopically \citep{Nugentetal1997,Kromeretal2010,WoosleyKasen2011}.
This drove a conclusion that a He shell thick enough to ignite and host a detonation would make a titanium, chromium, and nickel-rich outer layer that is not observed in SNe~Ia.
A few transients with the expected spectroscopic properties have been recently observed \citep{Jiangetal2017,Deetal2019,Polinetal2019}, demonstrating such explosions are possible, but must be quite rare compared to normal SNe~Ia.

In more recent multi-dimensional theoretical work \citep{Finketal2007,Finketal2010,Kromeretal2010,TownsleyMooreBildsten2012,MooreTownsleyBildsten2013}, it was found that if the He-layer burning is less complete than previously predicted, a viable scenario for normal SNe~Ia might be possible.
However, \cite{Kromeretal2010}, while demonstrating a very exciting possibility, stopped short of presenting a fully viable case.
Their best model was created by artificially removing the He shell ashes.
They were able to come close to this with a case that posited a 30\% enrichment of the He layer with C-O, but even this showed peculiar spectral features.
Working in one dimension, \citet{WoosleyKasen2011} confirmed that a He layer small enough would not lead to spectral abnormality.
Adding another essential element, \citet{Guillochonetal2010} proposed a dynamic ignition scenario for thin He shells during a double-degenerate merger.

Since those works, \cite{ShenMoore2014} found that by including a more complete nuclear reaction network in the simulation and a critical nitrogen isotope in the helium layer, the scenario becomes viable with helium shell masses $\sim 0.01 \, M_\odot$, which do not produce the iron-group material that plagued previous double-detonation simulations.
Here we follow up that work with a multi-dimensional simulation of the double-detonation scenario that produces a spectroscopically normal SN~Ia with a thin, modestly enriched He layer.

We choose a WD configuration that models a scenario in which the He shell is heated by a directly impacting accretion stream and mixes modestly with the outer edge of the core.
We use a 1.0 M$_\odot$ C-O core and a surface He layer with base $\rho=2\times 10^{5}$ g~cm$^{-3}$ and $T=5\times 10^{8}$ K, having a mass of  0.021~M$_{\odot}$.
The core has a uniform $T=3\times 10^{7}$ K and the temperature profile in the He shell declines outward adiabatically.
The composition of the core is, by mass, 0.4, 0.58, and 0.02 of $^{12}$C, $^{16}$O, and $^{22}$Ne respectively,
and the He shell is 0.891, 0.05, 0.009, and 0.05 of $^{4}$He, $^{12}$C, $^{14}$N, and $^{16}$O.
These shell abundances follow the fiducial case discussed in \citet{ShenMoore2014}.
The He detonation is ignited with a spherical hot spot placed on the symmetry axis at the base of the He layer.
The hot spot has a central temperature of $2\times 10^9$ K and a linear gradient falling to $8\times 10^{8}$ K at 200 km from the center.

After a brief discussion of the software used and nuclear reaction network chosen in section \ref{sec:method},
we present the results of our simulation in Section \ref{sec:results}.
We show the sequence of events in the double detonation, the nucleosynthetic products, and the resulting emergent radiation that would be observed from this explosion.
Finally, in section \ref{sec:conclusions}, we discuss the implications for the viability of double detonations as a scenario for spectroscopically normal SNe~Ia.

\section{Software and Reaction Network} 
\label{sec:method}

We use \texttt{FLASH}, version 4.3, to perform simulations of reactive, compressible fluid dynamics,
utilizing the split piecewise parabolic method for hydrodynamics and the ``new'' multipole gravity solver with a logarithmic radial grid.
The refinement strategy follows that in \citet{Townsleyetal2009}.
Mesh refinement is triggered by gradients in density, temperature, and fuel (He and/or C) abundance and energy-generating regions are refined to a maximum of 4~km resolution and non-energy-generating regions to 32~km.
\citet{Townsleyetal2009} showed that 16~km resolution in non-energy-generating regions was sufficient for hydrostatic balance and the ejecta-launching process to be converged.
We choose to use 32~km, as our progenitor WD is more than twice as large in radius compared to the Chandrasekhar-mass model tested by \citet{Townsleyetal2009}.
Here we use a Lagrangian scalar to distinguish between stellar material and the low-density ($10^{-3}$~g~cm$^{-3}$) material used to fill the remainder of the grid rather than a density threshold as used in \citet{Townsleyetal2009}.
In addition, the interface between the He shell and the C-O core is refined at 4 km resolution,
and the focus point of the inward shock, the site of the C detonation ignition, is refined ahead of the arrival of the shock.
Finally, once the burning has ceased, the maximum refinement is limited to maintain a resolution of between 512 and 1024 cells in radius out to the edge of the burned material as it expands.

We augment \texttt{FLASH}'s nuclear burning capability by coupling it to a 55-isotope nuclear network using the nuclear reaction module from \texttt{MESA}, version 9793.
The nuclear network is comprised of neutrons, $^1$H, $^4$He, $^{11}$B, $^{12-13}$C, $^{13-15}$N, $^{15-17}$O, $^{18}$F, $^{19-22}$Ne, $^{22-23}$Na, $^{23-26}$Mg, $^{25-27}$Al, $^{28-30}$Si, $^{29-31}$P, $^{31-33}$S, $^{33-35}$Cl, $^{36-39}$Ar, $^{39}$K, $^{40}$Ca, $^{43}$Sc, $^{44}$Ti, $^{47}$V, $^{48}$Cr, $^{51}$Mn, $^{52}$Fe, $^{56}$Fe, $^{55}$Co, and $^{56,58-59}$Ni.

This nuclear network is designed to obtain accurate energy release for high-temperature helium- and C/O-burning.
Specifically, the network yields energy releases within a few percent of a 495-isotope network at all relevant times for CNO-enriched helium at $\rho=10^6$ g cm$^{-3}$ and $T=10^{8.4-9.5}$ K,
and for C/O-rich material at $\rho = 10^{7.5}$ g cm$^{-3}$ and $T=10^{9.5-9.8}$ K.
\citet{Milesetal2019} demonstrated that a similar 41-isotope network also accurately reproduces the speed of steady-state, curved C-O detonations.
Note that only energetics are captured with this reduced network.  For accurate abundances, tracer particles must be used and post-processed with a much larger network, as we do in Section \ref{sec:ejecta}.
No burning limiter is used \citep{Kushnir_2013,Shen_18}, as yields from post-processing are fairly accurate for steady-state detonations \citep{Milesetal2019}.
The impact of a limiter on the simulation outcome deserves further study \citep{KatzZingale2019}.

\section{Results}
\label{sec:results}

\subsection{Explosion and Ejecta}
\label{sec:ejecta}

\begin{figure*}
\includegraphics[width=7.2in]{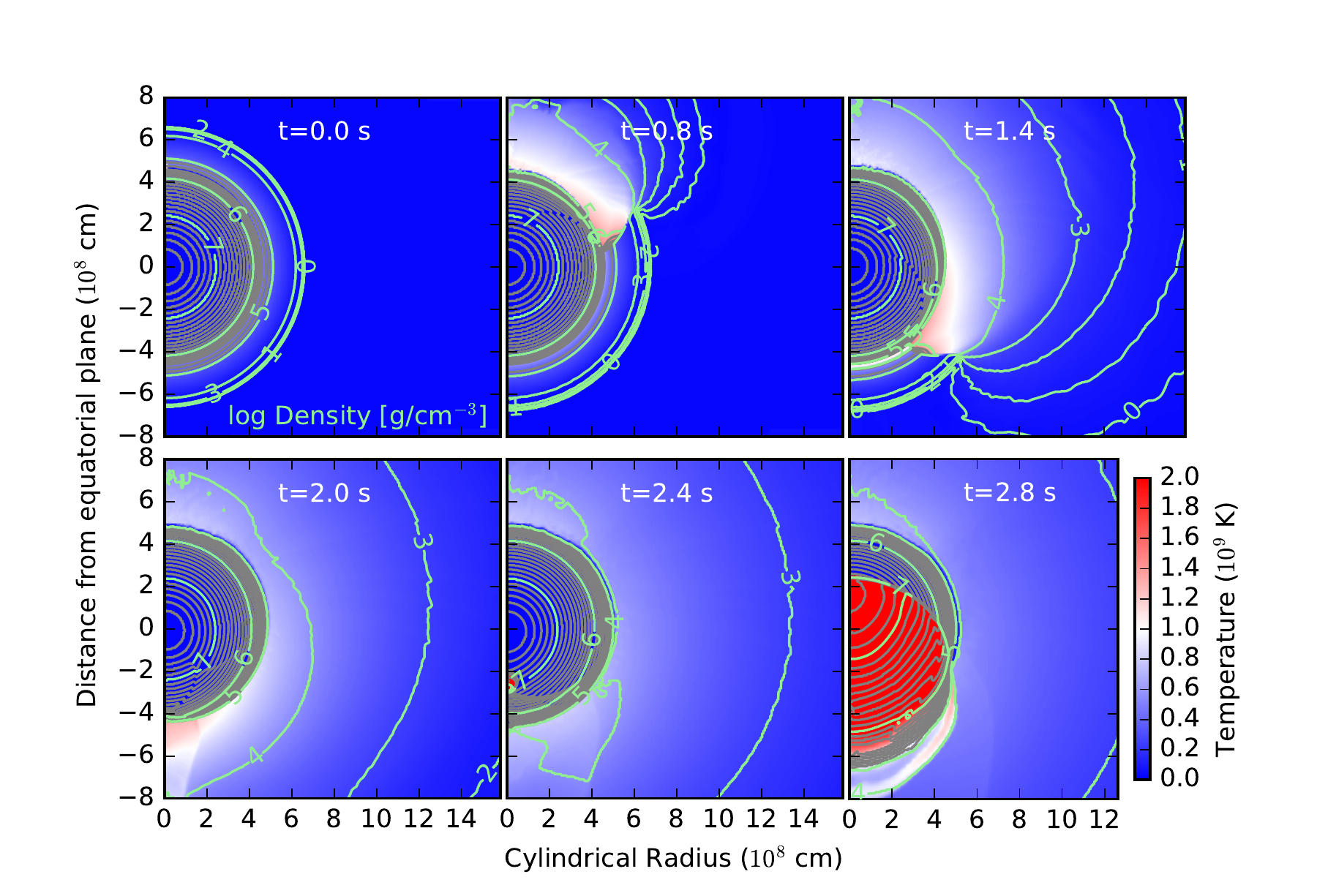}
\caption{
\label{fig:overview}
Time sequence of a helium shell double-detonation supernova.
Intermediate density contours are only shown above $10^5$~g~cm$^{-3}$, 10 per decade equally spaced logarithmically, in order to highlight the inward compression wave.
}
\end{figure*}

After being ignited in the helium shell, the explosion proceeds as expected for a double detonation with successive ignitions \citep{LivneGlasner1991,Finketal2010}.
A time sequence demonstrating an overview of the explosion appears in Figure \ref{fig:overview}.
The ignition is placed at the base of the helium shell as described in section \ref{sec:intro}.
The helium detonation propagates across the surface to the point opposite the ignition in just under 2~s.
Subsequently, the inward compression wave shed by the detonation is focused and ignites a carbon detonation just before 2.4~s about $2.5\times10^{8}$~cm from the center in the southern hemisphere.
When further refined to 1~km resolution, this ignition occurs about 12~km from the symmetry axis while the shock is moving inward toward the focus, as in \citet{ShenBildsten2014}.
Finally this detonation proceeds outward, incinerating the carbon inner region and passing out into the remnant helium shell just after 3~s.

\begin{figure*}
\includegraphics[width=7.2in]{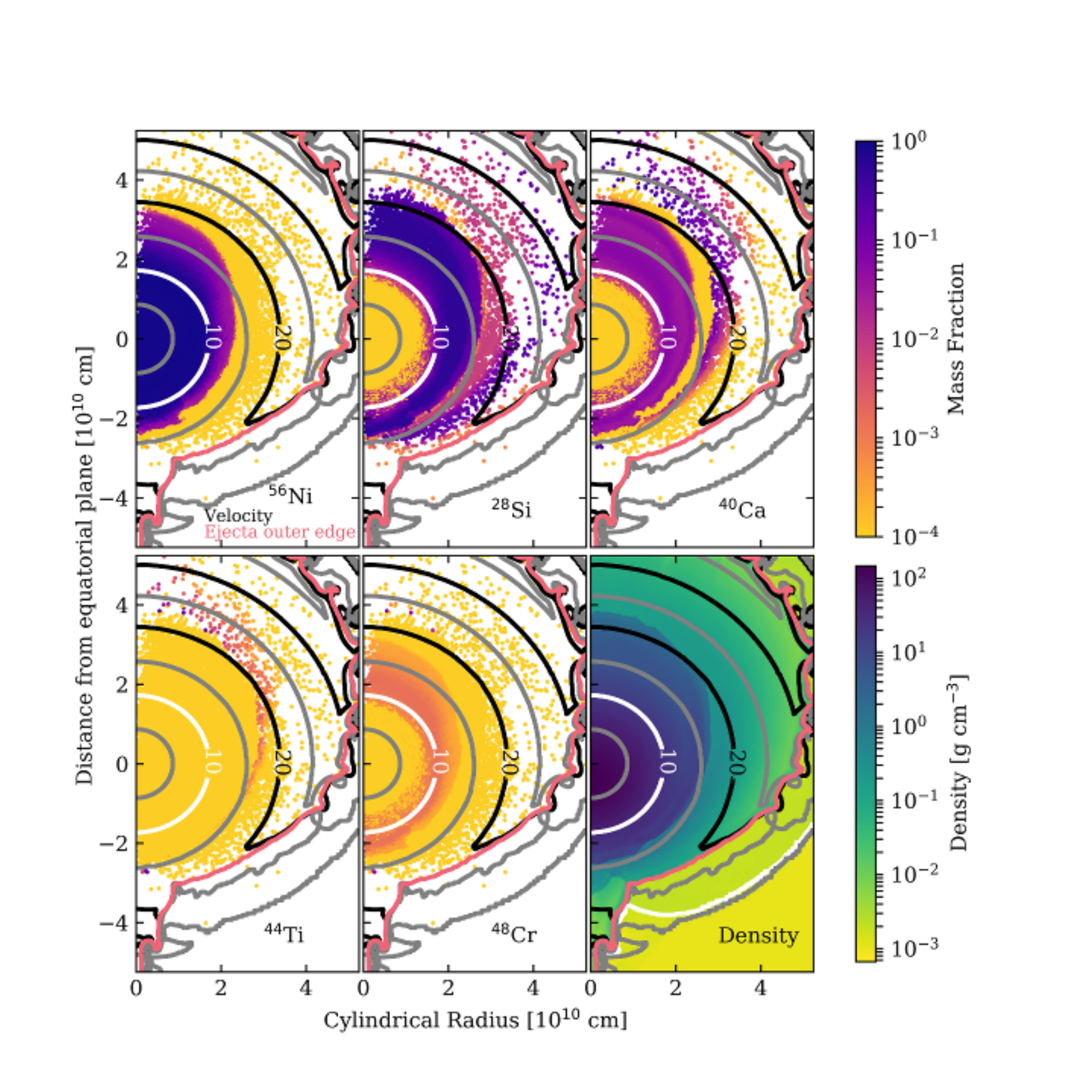}
\caption{
\label{fig:hfinal}
Ejecta 20~s after ignition, having entered the free expansion phase.  The
first five panels show the distribution of various species, indicated in the
lower right corner of each panel, at the location of post-processed Lagrangian tracers.  Overlaid are contours in total ejecta
velocity, labeled in units of $10^8$~cm~s$^{-1}$. The lower right panel shows the density distribution.  The red
contour indicates the edge of material that was part of the star at the
beginning of the simulation (see text).
Very little $^{44}$Ti, $^{48}$Cr, or heavier elements is made in the helium shell detonation.
}
\end{figure*}

After the detonation, the simulation is continued until 20~s, when the ejecta has settled into homologous expansion.
Final abundances are then determined by post-processing Lagrangian tracer histories with a 205-nuclide nuclear reaction network \citep{Milesetal2019}.
Initial abundances are for an initially solar metallicity progenitor in which C, N and O have been converted to $^{22}$Ne in the C/O core and to $^{14}$N in the He layer.
Other elements are solar, with elemental abundances from \citet{Asplund_metals}
and isotopic fractions from \citet{Lodders2003}, the convention used by MESA.
The distribution of several key species, including $^{56}$Ni, $^{28}$Si, $^{40}$Ca, $^{44}$Ti, and $^{48}$Cr, and the density are shown in Figure \ref{fig:hfinal}.
The red line indicates the boundary between material that was initially inside the star and low-density material, ``fluff,'' used to fill the hydrodynamic grid outside the star.
As the ejecta expands into the fluff, its outer edge is shocked and decelerated, with an outbound shock propagating outward into the fluff.
Anything outside the red line is discarded when post-processing.

Table \ref{yieldtable} gives the major nucleosynthetic yields of the explosion, for the He shell and overall, including high, intermediate, and low mass elements (HME, IME, LME) and other selected nuclides.
The final kinetic energy was $1.1\times 10^{51}$~erg and the total nuclear energy release was $1.3\times 10^{51}$~erg, as computed from either the hydrodynamic or post-processed tracer yields.
Total energy conservation was maintained to within a part in $10^4$.

\begin{table}
\caption{\label{yieldtable} Explosion Yields}
\begin{tabular}{lcc}
Species & He Shell (M$_\odot$) & Total (M$_\odot$)\\
\hline
$^{56}$Ni & $1.7\times 10^{-7}$  & 0.60 \\
HME ($Z\ge 21)$ - $^{56}{\rm Ni}$  & $3.0\times 10^{-5}$ & 0.039 \\
IME ($11\le Z\le20$)  & 0.0059  & 0.27 \\
LME ($Z\le 10$) & 0.015 & 0.092 \\
$^{12}$C & 0.00013 & 0.0021 \\
$^{28}$Si & 0.00076 & 0.14 \\
$^{40}$Ca & 0.00066 & 0.016 \\
$^{44}$Ti & $1.7\times 10^{-5}$ & $4.1\times 10^{-5}$ \\
$^{48}$Cr & $2.1\times 10^{-6}$ & $3.6\times 10^{-4}$ \\
\end{tabular}
\end{table}

Ti present during the photospheric phase will come predominantly from $^{48}$Cr via beta decay.
It is notable that very little $^{44}$Ti, $^{48}$Cr, and heavier elements are made in the helium shell detonation, less than made in the core.
A small amount is made near the ignition region and opposite that where the surface detonation converges.
This may be overemphasized by our ignition method.
The $^{56}$Ni distribution is mildly asymmetric, extending out to just below 14,000~km~s$^{-1}$ on the carbon ignition side and to about 19,000~km~s$^{-1}$ on the helium ignition side.
The Si and Ca distributions are more asymmetric, forming a double-layer structure with the outer layer arising from the helium shell ashes and the inner layer from the outer carbon-rich core.
Ca is synthesized at all latitudes, but more strongly in the highest.
These structures may result in the high-velocity features that are ubiquitous in SNe Ia; we leave a careful examination of this phenomenon to future work.

While equal-mass tracer particles ease computation of overall yields, they undersample the ejecta near the outer edge.
We therefore average the ejecta into 30$^{\circ}$ wedges in latitude for radiative post-processing.
Within each wedge, tracers are divided among 100 bins equally sized in radial velocity, with the velocity interval for each wedge determined by the maximum velocity reached by tracer particles in that region.
A maximum of 100 tracers were randomly selected within each bin for averaging.

The low-density, high-temperature outer layer bears some resemblance to the
structure in pulsational delayed detonation models
\citep{Dessartetal2014}, where the elevated temperature affords a
better match to the early time evolution of some observed SNe Ia.
There is a clear density discontinuity, whose
location in velocity varies with direction, at the interface between the
ashes of the C/O interior and the He shell.  This boundary lies at around
22,000~km~s$^{-1}$ near the ignition pole, falling to around
14,000~km~s$^{-1}$ in the direction near the lower pole.  There is also a
corresponding temperature discontinuity at this location, with the lower
density outer layer (He shell ashes) being hotter.
For the southern hemisphere wedges, we replace the portion of the profile beyond the position of the reverse shock (due to interaction with the fluff) with an extrapolated profile.

\subsection{Emergent Radiation}

\begin{figure*}
\includegraphics[width=\textwidth]{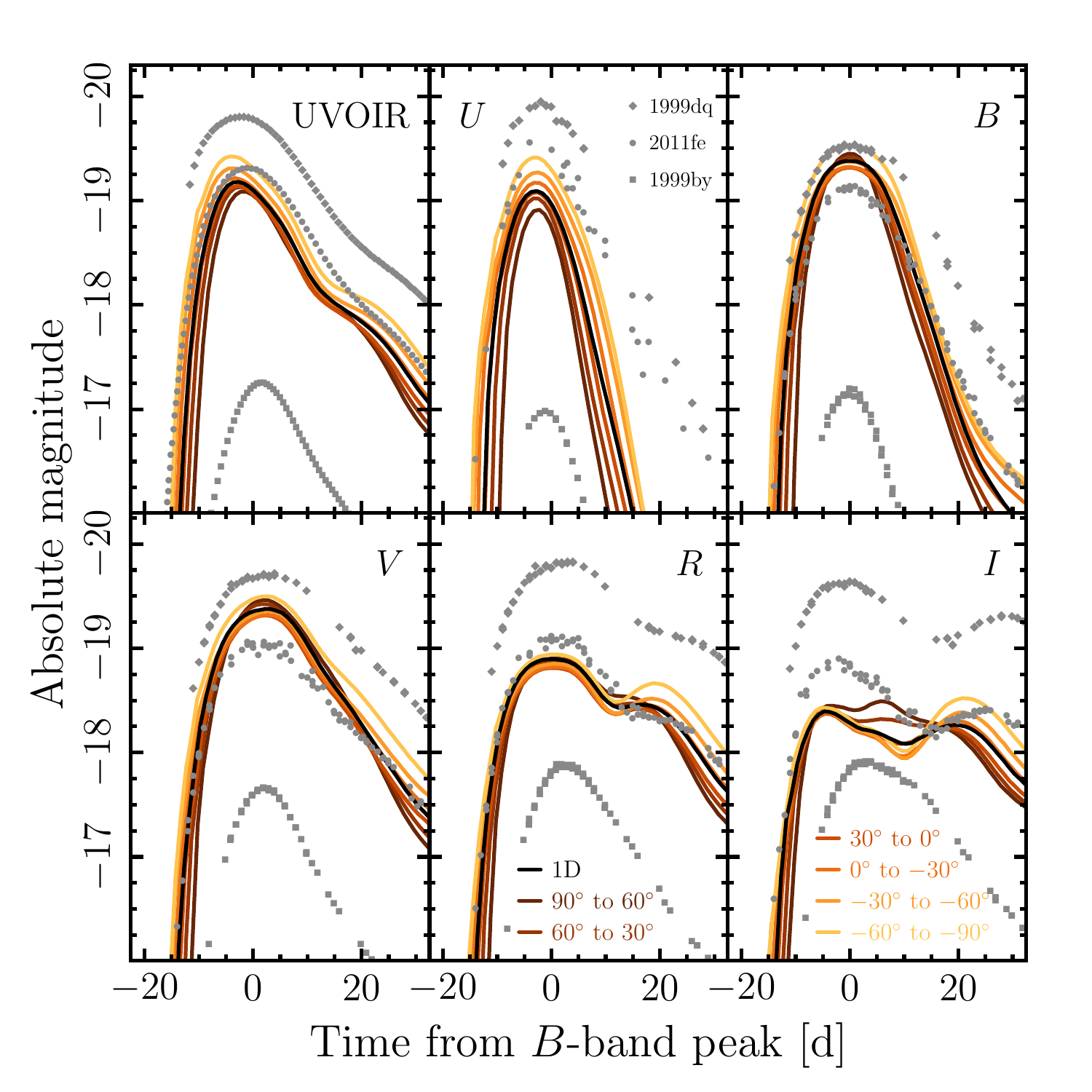}
\caption{\label{fig:lcs}
Time evolution of magnitude in various bands for different regions of ejecta.
The $1.0 \, M_\odot$ detonation model from \cite{Shen_18}, which has no He layer, is shown as black lines.
Gray points represent the sub-luminous SN 1999by, the normal SN 2011fe, and the over-luminous SN 1999dq.
}
\end{figure*}

Radiative transfer calculations were performed with \texttt{Sedona} \citep{kasen06a} for each of the wedges obtained.
Each is treated as the 1D abundance and density profile for an independent radiative transfer calculation.
Multi-band light curves are shown in Figure \ref{fig:lcs} along with that for a case from \citet{Shen_18} with the same WD mass (black lines).
The models from \citet{Shen_18} are centrally ignited detonations, computed in 1D, using a progenitor WD with uniform mass fractions of 0.495/0.495/0.01/0.001 $^{12}$C/$^{16}$O/$^{22}$Ne/$^{56}$Fe and no surface He layer.
The light curves from the mid-latitude wedges are fully consistent with the the case with no He layer in all passbands.
This demonstrates that the ashes produced by this thin, modestly enriched He shell have a negligible effect on the predicted colors of the supernova.
Importantly, we do not see the extremely red colors after maximum light obtained with thicker He shells \citep{Kromeretal2010,WoosleyKasen2011,Polinetal2019} caused by line blanketing.
However, as discussed in
\citet{Shen_18}, these \texttt{Sedona} light curves still decrease too quickly in the
\emph{U} and \emph{B} bands compared to observed light curves, which, based on comparison
to other work, we attribute to the local thermodynamic equilibrium (LTE) assumptions being made in this \texttt{Sedona}
calculation.

\begin{figure*}
\includegraphics[width=\textwidth]{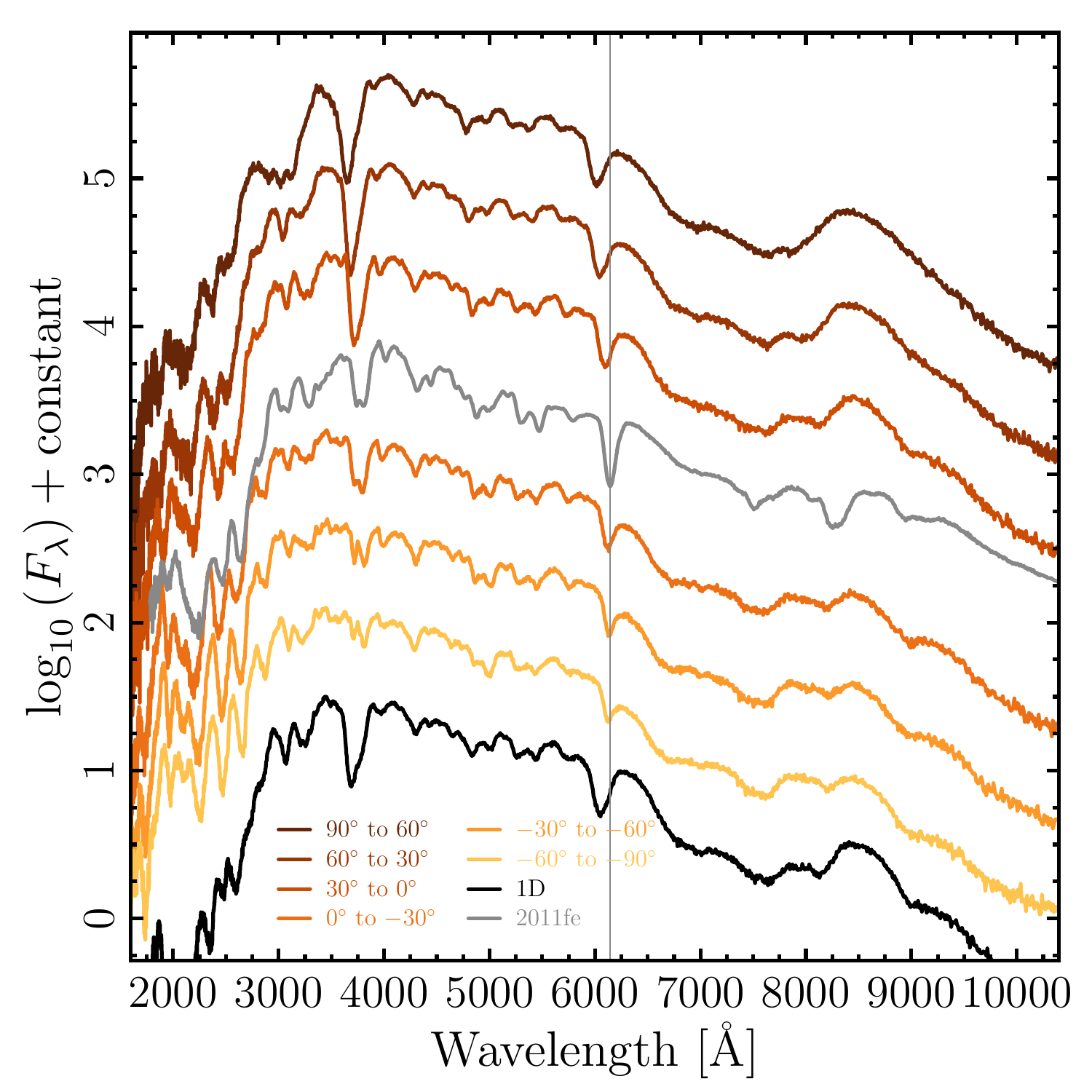}
\caption{\label{fig:spectra}
Maximum light spectra for ejecta representative of the six regions separated by angle from the equatorial plane.
The near-maximum-light spectrum of SN 2011fe is shown for comparison (gray), as well as the maximum-light spectrum from a 1.0~$M_\odot$ detonation model with no He layer (black) from \citet{Shen_18}.
The vertical line indicates the wavelength of the minimum of the Si {\sc ii} feature in SN 2011fe.
Even with the helium layer ashes, the spectra compares well to observations.
}
\end{figure*}

The maximum light spectra from this helium-ignited double-detonation
explosion are also a good match for spectroscopically normal
SNe Ia.
Figure \ref{fig:spectra} shows the spectra computed for each of the 30$^{\circ}$ wedges compared to the spectrum computed by \citet{Shen_18} (black) and the maximum light spectrum of SN 2011fe (gray; \citealt{mazzali14a}).
Spectra from the various wedges are offset by constants for display purposes.
We find that the spectra from mid-latitudes are again quite favorably comparable to both the centrally ignited detonation case with no He layer and to the spectrum of SN 2011fe.
Previous models have been too featureless near maximum light, without a strong enough Si {\sc ii} line \citep{Nugentetal1997}, related to the decreased velocity of the core imparted by the inward-going shock \citep{WoosleyKasen2011}.
The improvement is likely due to a combination of our model having no radioactivity in the outer layers, thus decreasing the ionization, and the production of Si in the He shell.

Features in our spectra are generally at slightly higher velocity than those
seen in SN 2011fe, but we have not attempted to fit the mass precisely.  The
blueshift of the trough of the major absorption features at 3800 and 6100 \AA\
is seen to increase at higher latitudes.
The ejecta reaches overall higher velocity in this direction due
to the core detonation being less curved, and therefore stronger, as it reaches the surface.
In the equatorial directions, more common viewing angles, the velocity mismatch compared to observations is minimal and is better than for the 1D case.
Ti absorption shortward of $\sim 3500$ \AA\ is more prominent in the higher latitude directions near the ignition point.
For wavelengths longer than $\sim 3500$ \AA, SN~2011fe shows good similarity to the spectrum from the $0^{\circ}$ to $-30^{\circ}$ wedge,
while for shorter wavelengths, the decline towards the UV in 2011fe is more comparable to the $30^{\circ}$ to $0^{\circ}$ and $60^{\circ}$ to $30^{\circ}$ wedges.

\section{Conclusions}
\label{sec:conclusions}

Following up on work showing that thin, highly C-O enriched He shells lead to double-detonations that nearly reproduce spectroscopically normal SNe~Ia \citep{Kromeretal2010}
and that a layer containing N, the product of the CNO cycle, can detonate at even lower densities with less complete burning \citep{ShenMoore2014},
we have simulated, in 2D, the double detonation of a 1.0~M$_\odot$ WD with a 0.02 M$_\odot$ He layer enriched with 5, 5, and 0.9\% $^{12}$C, $^{16}$O, and $^{14}$N,
finding that the resulting explosion produces a spectroscopically normal SN~Ia from a wide range of viewing angles.
This success provides a proof-of-concept that the dynamically driven double-degenerate double-detonation (D$^6$, \citealt{Shenetal2018D6}) scenario,
in which a thin He shell detonates early on in a WD-WD merger, can be a significant progenitor channel for a large fraction of SNe~Ia.

While our spectra compare well with normal SNe~Ia, the decline rate of our $B$ band light-curve is too fast.
This was also true for the centrally ignited detonations without helium layers computed by \citet{Shen_18} using the same radiative transfer as the present work.
We believe this is a remaining uncertainty in the radiative transfer that will be addressed in future work.
This conclusion is based on the fact that the decline rate is a better match for observations for similar models computed by \citet{Simetal2010} and \citet{blon17a}.
These works used methods intended to capture non-LTE effects to varying degrees in ways that should be more realistic.
The case shown here also motivates work to determine the range of He shell sizes, enrichments, and thermal states that are allowed by observations,
as well as further study of the dynamically driven ignition mechanism in mergers of WD-WD binaries.

\acknowledgements

B.J.M., D.M.T., and K.J.S. received support from the NASA Astrophysics Theory Program (NNX17AG28G).
We thank Carla Fr\"olich for supporting B.J.M.'s participation.
Portions of this work were supported by the United States Department of Energy, under an Early Career Award (Grant No. SC0010263),
by the Office of Science, Office of Nuclear Physics, award DE-FG02-02ER41216, awards DE-AC02-05CH11231 and DE-SC0017616 (D.K.),
by SciDAC award DE-SC0018297 (D.K.),
by the Research Corporation for Science Advancement under Cottrell Scholar Awards,
and by the Gordon and Betty Moore Foundation through Grant GBMF5076 (D.K.).
We utilised the National Energy Research Scientific Computing Center, a DOE Office of Science User Facility supported by the Office of Science of the U.S. Department of Energy (Contract No. DE-AC02-05CH11231).


\software{\texttt{FLASH} (\citealt{Fryxelletal2000,Dubeyetal2009,Dubeyetal2013,Dubeyetal2014}, \url{flash.uchicago.edu}),
\texttt{MESA} (\citealt{paxton_2011,paxton_2013,paxton_2015,paxton_2018}, \url{mesa.sourceforge.net}),
\texttt{Sedona} \citep{kasen06a}},
\texttt{yt} (\url{yt-project.org})

\bibliography{master,timmes_master,townsley_master,detonations,ken}

\end{document}